\documentclass[screen,sigconf]{acmart}
\AtBeginDocument{%
  \providecommand\BibTeX{{%
    \normalfont B\kern-0.5em{\scshape i\kern-0.25em b}\kern-0.8em\TeX}}}

\acmYear{2021}\copyrightyear{2021}
\setcopyright{rightsretained}
\acmConference[ACM CHIL '21]{ACM Conference on Health, Inference, and Learning}{April 8--10, 2021}{Virtual Event, USA}
\acmBooktitle{ACM Conference on Health, Inference, and Learning (ACM CHIL '21), April 8--10, 2021, Virtual Event, USA}
\acmPrice{}
\acmDOI{10.1145/3450439.3451862}
\acmISBN{978-1-4503-8359-2/21/04}

\usepackage{makecell}
\begin{document}

\title{VisualCheXbert: Addressing the Discrepancy Between Radiology Report Labels and Image Labels}

\author{Saahil Jain}
\authornote{Equal Contribution}
\email{saahil.jain@cs.stanford.edu}
\affiliation{%
  \institution{Stanford University}
    \country{USA}
}

\author{Akshay Smit}
\authornotemark[1]
\email{akshaysm@stanford.edu}
\affiliation{%
  \institution{Stanford University}
\country{USA}
}

\author{Steven QH Truong}
\affiliation{%
  \institution{VinBrain}
\country{Vietnam}
}

\author{Chanh DT Nguyen}
\author{Minh-Thanh Huynh}
\affiliation{%
  \institution{VinBrain}
\country{Vietnam}
}

\author{Mudit Jain}
\affiliation{\institution{}
\country{USA}}

\author{Victoria A. Young}
\affiliation{%
  \institution{Stanford University}
\country{USA}
}

\author{Andrew Y. Ng}
\email{ang@cs.stanford.edu}
\affiliation{%
  \institution{Stanford University}
\country{USA}
}
 
\author{Matthew P. Lungren}
\authornote{Equal Contribution}
\email{mlungren@stanford.edu}
\affiliation{%
  \institution{Stanford University}
\country{USA}
}

\author{Pranav Rajpurkar}
\authornotemark[2]
\email{pranavsr@cs.stanford.edu}
\affiliation{%
  \institution{Stanford University}
\country{USA}
}
\renewcommand{\shortauthors}{Jain \& Smit et al.}

\begin{abstract}
Automatic extraction of medical conditions from free-text radiology reports is critical for supervising computer vision models to interpret medical images. In this work, we show that radiologists labeling reports significantly disagree with radiologists labeling corresponding chest X-ray images, which reduces the quality of report labels as proxies for image labels. We develop and evaluate methods to produce labels from radiology reports that have better agreement with radiologists labeling images. Our best performing method, called VisualCheXbert, uses a biomedically-pretrained BERT model to directly map from a radiology report to the image labels, with a supervisory signal determined by a computer vision model trained to detect medical conditions from chest X-ray images. We find that VisualCheXbert outperforms an approach using an existing radiology report labeler by an average F1 score of 0.14 (95\% CI 0.12, 0.17). We also find that VisualCheXbert better agrees with radiologists labeling chest X-ray images than do radiologists labeling the corresponding radiology reports by an average F1 score across several medical conditions of between 0.12 (95\% CI 0.09, 0.15) and 0.21 (95\% CI 0.18, 0.24).
\end{abstract}



\begin{teaserfigure}
\centering
  \includegraphics[width=0.7\textwidth]{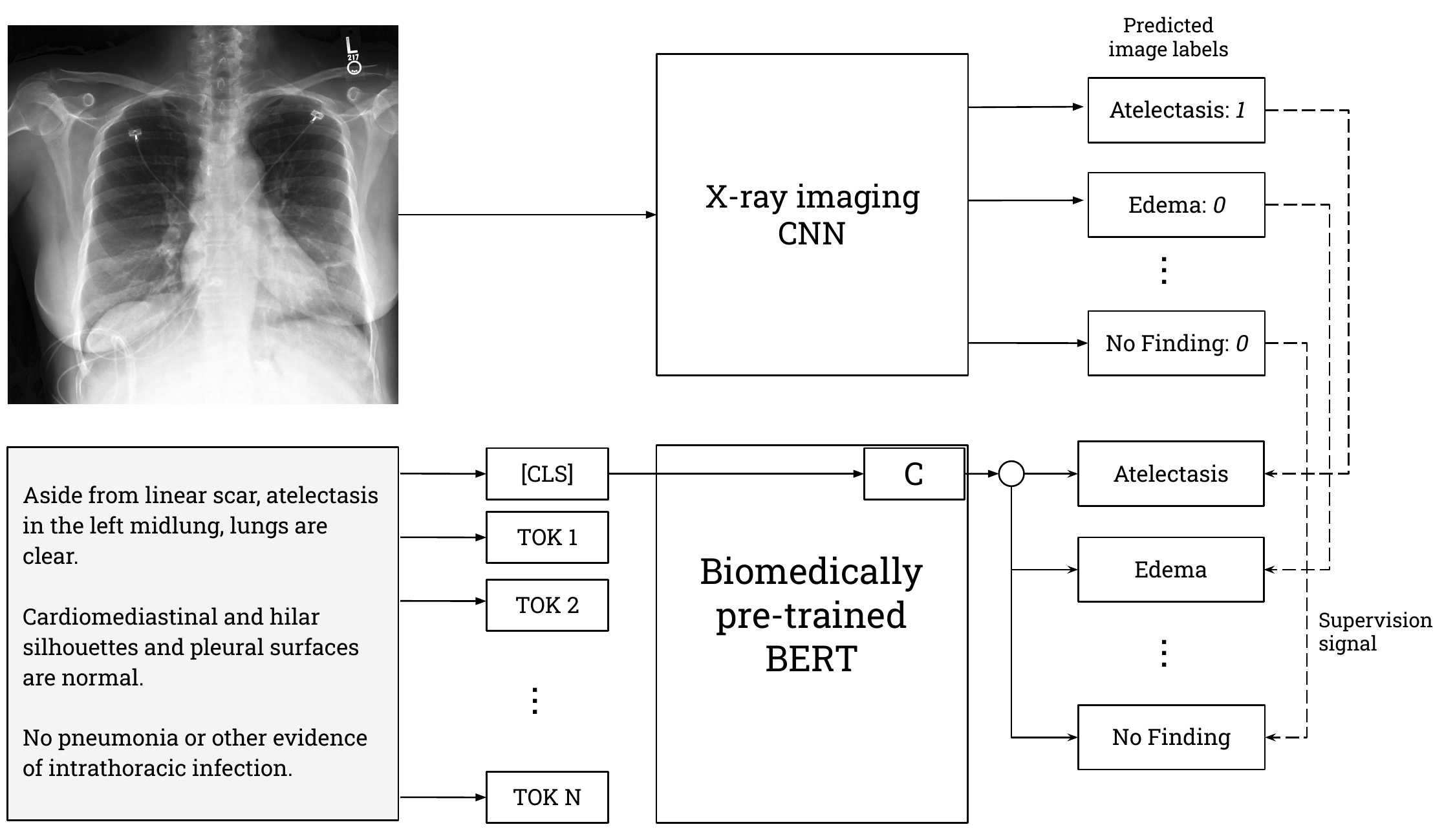}
  \caption{The VisualCheXbert training procedure. VisualCheXbert uses a biomedically-pretrained BERT model to directly map from a radiology report to the labels obtained by a radiologist interpreting the associated X-ray image. The training procedure for VisualCheXbert is supervised by a computer vision model trained to detect medical conditions from chest X-ray images.}
  \Description{VisualCheXbert architecture}
  \label{fig:teaser}
\end{teaserfigure}


\maketitle

\section{Introduction}
Because manually annotating a large number of medical images is costly \cite{nguyen2021vindrcxr, Bustos_2020, Abramoff:2016aa, Gulshan:2016aa, doi:10.1148, covidnet, Demner-Fushman:2016aa}, an appealing solution is the use of automatic labelers to extract labels from medical text reports that accompany the images. On the task of chest X-ray interpretation, high-performing vision models have been successfully trained \cite{rajpurkar2020chexpedition, pham2020interpreting, ye2020weakly, rajpurkar2020chexphotogenic, tang2020nature, rajpurkar2020chexaid} on large, publicly available chest X-ray datasets \cite{irvin2019chexpert, johnson2019mimiccxrjpg, wang2017chestx, phillips2020chexphoto} labeled by automated radiology report labelers \cite{irvin2019chexpert, peng2017negbio, mcdermott2020chexpert, smit2020chexbert}. However, training these vision models on labels obtained from reports assumes that the report labels are good proxies for image labels. Prior work has found that report labels may not accurately reflect the visual content of medical images \cite{oakdenrayner2019exploring, olatunji2019caveats, tang2020data}.

We investigate this assumption in the setting of automated chest X-ray labeling and develop methods to produce labels from radiology reports that better agree with radiologists labeling the corresponding X-ray images. Our primary contributions are:
\begin{enumerate}
    \item We quantify the agreement between radiologists labeling reports and radiologists labeling images across several medical conditions. We find that there is significant disagreement between board-certified radiologists when labeling a chest X-ray image and when labeling the corresponding radiology report.
    \item Upon board-certified radiologist review of examples of disagreements between radiologists labeling reports and radiologists labeling images, we find various reasons for disagreement related to (a) label hierarchy relationships, (b) access to clinical history, (c) the use of the \textit{Impression} and \textit{Findings} section of radiology reports, and (d) the inherent noise of the labeling task.
    \item We find many significant relationships between presence of conditions labeled using reports and presence of conditions labeled using images. We report and clinically interpret various radiology report labels that increase (or decrease) the odds of particular conditions in an image with statistical significance.
    \item We learn to map textual radiology reports directly to the X-ray image labels. Our best performing method, called \textit{VisualCheXbert}, uses a biomedically-pretrained BERT model to directly map from a radiology report to the image labels. We find that VisualCheXbert better agrees with radiologists labeling chest X-ray images than do radiologists labeling the corresponding radiology reports by an average F1 score across several medical conditions of between 0.12  (95\% CI 0.09, 0.15) and 0.21 (95\% CI 0.18, 0.24). We also find that VisualCheXbert outperforms an approach using the CheXpert radiology report labeler \cite{irvin2019chexpert} by an average F1 score of 0.14 (95\% CI 0.12, 0.17).
\end{enumerate}

We expect that our methods of addressing the discrepancy between medical report labels and image labels are broadly useful across the medical domain and may facilitate the development of improved medical imaging models.

\section{Data}

We made use of two large publicly available datasets of chest X-rays: CheXpert \cite{irvin2019chexpert} and MIMIC-CXR \cite{johnson2019mimiccxrjpg}. For both datasets, we use the \textit{Impression} section of the radiology reports, which summarizes the key findings in the radiographic study. Each of the X-rays in these datasets was labeled for 14 commonly occurring medical conditions. CheXpert consists of 224,316 chest radiographs, with labels generated from the corresponding radiology report impression by the automatic, rules-based CheXpert labeler. Given a radiology report impression as input, the CheXpert labeler labels each medical condition (except ``No Finding'') as ``positive'', ``negative'', ``uncertain'' or ``blank''. A ``blank'' label is produced by the CheXpert labeler if the condition was not mentioned at all in the report impression. If the condition was mentioned but its presence was negated, a ``negative'' label is produced. If the condition was mentioned but its presence was uncertain, an ``uncertain'' label is produced. For ``No Finding'', the CheXpert labeler only produces ``positive'' or ``blank'' labels. ``No Finding'' is only labeled as ``positive'' if no medical abnormality whatsoever was mentioned in the report impression. The MIMIC-CXR dataset consists of 377,110 chest X-rays and their corresponding radiology reports, and it has also been labeled by the CheXpert labeler. 

The CheXpert dataset contains a separate set of 200 chest X-ray studies called the ``CheXpert validation set'' and another set of 500 chest X-ray studies called the ``CheXpert test set''. The CheXpert validation set is labeled by the majority vote of 3 board-certified radiologists examining the X-ray images and labeling each of the 14 conditions as ``positive'' or ``negative'', similar to the image ground truth on the CheXpert test set, which is described below. No radiologist report labels are obtained for the validation set. 

The CheXpert test set, which was collected by \citet{irvin2019chexpert}, is labeled by radiologists in two distinct ways:

\paragraph{Image ground truth} 5 board-certified radiologists looked at each X-ray image and labeled each of the 14 conditions as ``positive'' or ``negative''. The final label is their majority vote. These radiologists only observed the X-ray images and did not have access to the radiology report or patients' historical records at the time of image labeling.

\paragraph{Radiologist report labels} A board-certified radiologist looked at each radiology report impression corresponding to the X-rays and labeled each of the 14 conditions as being ``positive'', ``negative'', ``uncertain'', or ``blank''. This radiologist did not observe any X-ray images. A condition was labeled as ``blank'' if it was not at all mentioned in the report impression. If the condition was mentioned but its presence in the chest X-ray was negated, then the condition was labeled as ``negative''. If the condition was mentioned but its presence was uncertain, it was labeled as ``uncertain''.

\begin{table}[t!]
\centering
\caption{\label{tab:report_image_agreement} Agreement between radiologists looking at reports and radiologists looking at the corresponding X-ray images. The high and low scores are obtained by mapping uncertain labels in the radiologist report labels to the image ground truth labels and the opposite of the image ground truth labels respectively.}
\resizebox{\columnwidth}{!}{%
\begin{tabular}{lrrrr}
\textbf{\begin{tabular}[c]{@{}l@{}}Condition\\ (n = \# positive)\end{tabular}} & \multicolumn{1}{l}{\textbf{Low F1}} & \multicolumn{1}{l}{\textbf{High F1}} & \multicolumn{1}{l}{\textbf{Low Kappa}} & \multicolumn{1}{l}{\textbf{High Kappa}} \\ \hline
Atelectasis (n=153)                                                            & 0.230                               & 0.595                                & -0.014                                 & 0.457                                   \\
Cardiomegaly (n=151)                                                           & 0.422                               & 0.463                                & 0.290                                  & 0.344                                   \\
Edema (n=78)                                                                   & 0.453                               & 0.581                                & 0.335                                  & 0.492                                   \\
Pleural Effusion (n=104)                                                       & 0.638                               & 0.710                                & 0.511                                  & 0.613                                   \\
Enlarged Cardiom. (n=253)                                             & 0.089                               & 0.208                                & -0.053                                 & 0.097                                   \\
Lung Opacity (n=264)                                                           & 0.683                               & 0.686                                & 0.401                                  & 0.405                                   \\
Support Devices (n=261)                                                        & 0.863                               & 0.863                                & 0.737                                  & 0.737                                   \\
No Finding (n=62)                                                              & 0.381                               & 0.381                                & 0.292                                  & 0.292                                   \\ \hline
Average                                                                        & 0.470                               & 0.561                                & 0.312                                  & 0.430                                   \\
Weighted Average                                                               & 0.492                               & 0.575                                & 0.320                                  & 0.427                        
\end{tabular}%
}
\end{table}



\section{Evaluation}
We only evaluate our models on medical conditions for which at least 50 out of the 500 chest X-ray studies in the CheXpert test set were marked positive by the radiologists labeling the X-ray images (image ground truth). These conditions, which we refer to as the \textit{evaluation conditions}, are: Atelectasis, Cardiomegaly, Edema, Pleural Effusion, Enlarged Cardiomediastinum, Lung Opacity, Support Devices, and No Finding. We evaluate models using the average and weighted average of the F1 score across conditions on the CheXpert test set with the image ground truth. To compute the weighted average, each condition is weighted by the portion of positive labels for that condition in the CheXpert test set. 


\begin{table*}[ht]
\centering
\caption{\label{tab:report_examples} Clinical explanations of disagreements between radiologists looking at reports and radiologists looking at images on the CheXpert test set. Given access to the X-ray image, the full radiology report, the radiology report impression, the radiology report labels, and the image ground truth, a board-certified radiologist explained disagreements between radiologist report labels and the image ground truth. We show select examples with explanations in this table.}
\resizebox{\textwidth}{!}{%
\begin{tabular}{|l|l|}
\hline
\textbf{Report Impression and Labels}                                                                                                                                                                                                                                                                                                                                                                                                                                                                                                          & \textbf{Clinical Explanation}                                                                                                                                                                                                                                                                                                                                                                                                                                                                                  \\ \hline
\begin{tabular}[c]{@{}l@{}}1. single ap upright view of the chest showing a mildly increased\\ opacity at the left lung base that could represent atelectasis\\ versus consolidation.\\ \\ \makecell[r]{\textit{Cardiomegaly}\\ Radiologist Report Label: Negative\\ Image Ground Truth: Positive}

\end{tabular}                                                                                                                                                                                                                                  & \begin{tabular}[c]{@{}l@{}}The radiologist looking at the report marks Cardiomegaly as \\ negative as it is not mentioned in the report.  Since the image is a \\ Intensive Care Unit (ICU) film and cardiomegaly is not a clinically \\ relevant condition for the population selected for in ICU films, the \\ presence of cardiomegaly was never mentioned in the report, \\ resulting in the discrepancy between radiologists looking at the \\ report and radiologists looking at the image.\end{tabular} \\ \hline
\begin{tabular}[c]{@{}l@{}}1. pulmonary vascular congestion. left lower lobe opacity comp-\\-atible with atelectasis and/or consolidation.\\ \\ \makecell[r]{\textit{Cardiomegaly}\\ Radiologist Report Label: Negative\\ Image Ground Truth: Positive}\end{tabular}                                                                                                                                                                                                                                                                               & \begin{tabular}[c]{@{}l@{}}Although cardiomegaly was mentioned in the radiology report \\ "Findings" section, cardiomegaly was not mentioned in the report \\ "Impression". Since the radiologist looking at the report only had \\ access to the "Impression" section, they labeled Cardiomegaly as \\ negative when it was actually present in the image.\end{tabular}                                                                                                                                           \\ \hline
\begin{tabular}[c]{@{}l@{}}1.  decreased pulmonary edema.  stable bilateral pleural effusions \\ and bibasilar atelectasis.\\ \\ \makecell[r]{\textit{Edema}\\ Radiologist Report Label: Positive\\ Image Ground Truth: Negative}\end{tabular}                                                                                                                                                                                                                                                                                                   & \begin{tabular}[c]{@{}l@{}}The phrase "decreased pulmonary edema" shows that the radiologist \\ writing the report had relevant clinical context, as the edema has \\ "decreased" compared to a previous report or image. However, the \\ radiologist looking at the image does not have this clinical context, \\ resulting in a discrepancy.\end{tabular}                                                                                                                                                     \\ \hline
\begin{tabular}[c]{@{}l@{}}1.  single frontal radiograph of the chest is limited secondary to \\ poor inspiration and rotation.  \\ 2.  cardiac silhouette is partially obscured secondary to rotation.  \\ lungs demonstrate bibasilar opacities, likely reflecting \\ atelectasis. possible small right pleural effusion. no pneumotho- \\ -rax.  \\ 3.  visualized osseous structures and soft tissues unremarkable.\\ \\ \makecell[r]{\textit{Pleural Effusion}\\ Radiologist Report Label: Positive\\ Image Ground Truth: Negative}\end{tabular} 
& \begin{tabular}[b]{@{}l@{}}The phrase "possible small right pleural effusion" indicates the \\ uncertainty regarding the presence of pleural effusion. This natural \\ uncertainty may explain the disagreement between radiologists \\ looking at the image and radiologists looking at the report. On \\ review, it was noted that pleural effusion was borderline in this \\ example.\end{tabular}                                                                                       \\ \hline
\begin{tabular}[c]{@{}l@{}}1. crowding of the pulmonary vasculature. cannot exclude mild\\ interstitial pulmonary edema.\\ 2. no focal air space consolidation. the cardiomediastinal silhou-\\-ette appears grossly within normal limits.\\ \\ \makecell[r]{\textit{Pleural Effusion}\\ Radiologist Report Label: Negative\\ Image Ground Truth: Positive}\end{tabular}                                                                                                                                                                           & \begin{tabular}[b]{@{}l@{}} Upon review by a board-certified radiologist, there was an error in \\ the radiology report, which did not mention the presence of pleural \\ effusion. The error in the report itself may explain the disagreement \\ between the image and report labels.\end{tabular}                                                                                                                                                                                                                                            \\ \hline
\end{tabular}}
\end{table*}

\section{Experiments}

\subsection{Do radiologists labeling reports agree with radiologists labeling X-ray images?}

We first investigate the extent of the disagreement between board-certified radiologists when labeling a chest X-ray image and when labeling the corresponding radiology report. 

\subsubsection{Method}
We compute the level of agreement between radiologists labeling X-ray images and radiologists labeling the corresponding radiology reports on the CheXpert test set. The CheXpert test set contains a set of labels from radiologists labeling X-ray images as well as another set of labels from radiologists labeling the corresponding radiology reports. Using the labels from X-ray images as the ground truth, we compute Cohen's Kappa \cite{cohenkappa} as well as the F1 score to measure the agreement between these two sets of labels. To compare the radiologist report labels to the image ground truth labels, we convert the radiologist report labels to binary labels as follows. We map the blank labels produced for the radiology report to negative labels. We map uncertain labels to either the image ground truth label or the opposite of the image ground truth label, and we record the results for both these strategies to obtain ``Low F1'', ``High F1'', ``Low Kappa'', and ``High Kappa'' scores. The low and high scores represent the most pessimistic and optimistic mapping of the uncertainty labels.

\subsubsection{Results}
We find that there is significant disagreement, which is indicated by low Kappa and F1 scores for almost all conditions evaluated. For example, Enlarged Cardiomediastinum and No Finding have a relatively small ``High Kappa'' score of 0.097 and 0.292 and a ``High F1'' score of 0.208 and 0.381, indicating high levels of disagreement even when assuming the most optimistic mapping of the uncertainty labels. Atelectasis, Cardiomegaly, Edema, Pleural Effusion, and Lung Opacity also have a low ``High Kappa'' score of 0.457, 0.344, 0.492, 0.613, and 0.405 respectively and a ``High F1'' score of 0.595, 0.463, 0.581, 0.710, and 0.686 respectively. Support Devices has the highest Kappa score, with a ``High Kappa'' of 0.737, and the highest F1 score, with a ``High F1'' of 0.863. The average Kappa score is between 0.312 and 0.430, and the average F1 score is between 0.470 and 0.561. The low and high F1 / Kappa scores for the evaluation conditions are shown in Table \ref{tab:report_image_agreement}.

\subsection{Why do radiologists labeling reports disagree with radiologists labeling X-ray images?}
We investigate why there is disagreement between board-certified radiologists when labeling a chest X-ray image and when labeling the corresponding radiology report.

\subsubsection{Method}
A board-certified radiologist was given access to the chest X-ray image, the full radiology report, the radiology report impression section, the image ground truth across all conditions, and the radiologist report labels across all conditions for each of the 500 examples in the CheXpert test set. The radiologist then explained examples where radiologists labeling reports disagree with radiologists labeling X-ray images. We also calculated the counts of disagreements between radiologists labeling reports and radiologists labeling X-ray images for each condition on the CheXpert test set. A board-certified radiologist explained why there were large numbers of disagreements on certain conditions.

\subsubsection{Results}
We find various reasons why radiologists labeling reports might disagree with radiologists labeling images. First, there is a difference between the setup of the report labeling and image labeling tasks related to the label hierarchy. On the report labeling task on the CheXpert test set, radiologists were instructed to label only the most specific condition as positive and leave parent conditions blank. For example, although Lung Opacity is a parent condition of Edema, a radiologist marking a report as positive for Edema would leave Lung Opacity blank. Blank report labels are typically mapped to negative image labels. However, radiologists labeling images label each condition as positive or negative independent of the presence of other conditions. Second, radiologists labeling reports have access to clinical report history, which biases radiologists towards reporting certain conditions in reports while a radiologist labeling the image may not observe the condition on the image. \citet{busby2018bias} explain biases from clinical history in terms of framing bias, where the presentation of the clinical history can lead to different diagnostic conclusions, and attribution bias, where information in the clinical history can lead to different diagnostic conclusions. Third, radiologists labeling reports were only given access to the report impression section when labeling the CheXpert test set. Sometimes, conditions are mentioned in the \textit{Findings} section of the report but not mentioned in the \textit{Impression} section. This results in more negative labels when radiologists looked at reports. For chest CT scan reports, \citet{gershanik2011critical} also find that a condition mentioned in the \textit{Findings} section is not always mentioned in the \textit{Impression} section of the report. Fourth, labeling images and reports is inherently noisy to a certain extent, resulting in disagreement. Drivers of noise include mistakes on the part of radiologists labeling reports and radiologists labeling images, uncertainty regarding the presence of a condition based on an image or report, and different thresholds for diagnosing conditions as positive among radiologists. \citet{brady2012discrepancy} describe additional factors that contribute to discrepancies in radiologist interpretations, including radiologist specific causes of error like under reading as well as system issues like excess workload. \citet{wood2020labelling}, in their analysis on MRI neuroradiology reports, also note factors, such as a difference in observations within reports depending on the referrer, that likewise result in discrepancies.

Next, we explain the counts of the largest disagreements between radiologists labeling reports and radiologists labeling images. Out of the 500 examples on the CheXpert test set, there were 223 examples where the image was labeled positive while the report was labeled negative for Enlarged Cardiomediastinum. We hypothesize that this results from the difference in the task setup related to the label hierarchy. Since Enlarged Cardiomediastinum is a parent condition of Cardiomegaly, radiologists labeling reports were instructed to leave Enlarged Cardiomediastinum blank if they labeled Cardiomegaly positive. There were 101 examples where the image was labeled positive while the report was labeled negative for Cardiomegaly. Diagnosis of cardiomegaly on chest radiographs can depend on patient positioning and clinical history. Further, particularly in the ICU setting in which multiple consecutive radiographs are taken, cardiomegaly is not consistently described in the report even when present unless a clinically significant change is observed (i.e. pericardial effusion). There were 100 examples where the image was labeled positive while the report was labeled negative for Lung Opacity. We hypothesize that this results from the difference in task setup related to label hierarchy, as Lung Opacity is a parent condition. Further, particularly in the setting of atelectasis, lung opacity may not have risen to clinical relevance for the reporting radiologist despite being seen on the isolated imaging task. There were 65 examples where the image was labeled negative while the report was labeled positive for Pleural Effusion. We hypothesize that this partially results from both the variant thresholds for diagnosis of pleural effusion among radiologists and the clinical setting in which the reporting radiologist has access to prior films. It was common to see the report state "decreased" or "trace residual" effusion due to the context of prior imaging on that patient. However, in the isolated image labeling task, the perceived likelihood of the condition fell below the threshold of a board-certified radiologist. There were 49 examples where the image was labeled negative, while the report was labeled positive for Edema. Similar to the effusion example, clinical context and prior imaging played a role in these discrepancies as, again, diagnoses were carried forward from prior studies and language such as "some residual" or "nearly resolved" in the report were used to indicate the presence of edema based on the clinical context. However, when labeling the corresponding image in isolation, the presence of edema fell below the threshold of a board-certified radiologist. Table \ref{tab:report_examples} contains specific examples of these disagreements with clinical explanations. Table \ref{tab:report_image_disagreement_counts} shows the counts of disagreements between radiologists labeling reports and radiologists labeling images by condition. 

\begin{table}[ht]
\caption{\label{tab:report_image_disagreement_counts} Counts of disagreements by condition between radiologists labeling reports and radiologists labeling the corresponding X-ray images on the CheXpert test set. The first column reports the number of times the image ground truth was positive, while the radiologist report label was negative. The second column reports the number of times the image ground truth was negative, while the radiologist report label was positive.}
\resizebox{\columnwidth}{!}{%
\begin{tabular}{|l|r|r|}
\hline
\textbf{Condition}         & \multicolumn{1}{l|}{\textbf{\begin{tabular}[c]{@{}l@{}}Positive on Image\\ Negative on Report\end{tabular}}} & \multicolumn{1}{l|}{\textbf{\begin{tabular}[c]{@{}l@{}}Negative on Image\\ Positive on Report\end{tabular}}} \\ \hline
No Finding                 & 38                                                                                                           & 40                                                                                                           \\ \hline
Enlarged Cardiom. & 223                                                                                                          & 5                                                                                                            \\ \hline
Cardiomegaly               & 101                                                                                                          & 15                                                                                                           \\ \hline
Lung Opacity               & 100                                                                                                          & 50                                                                                                           \\ \hline
Lung Lesion                & 2                                                                                                            & 12                                                                                                           \\ \hline
Edema                      & 26                                                                                                           & 49                                                                                                           \\ \hline
Consolidation              & 16                                                                                                           & 17                                                                                                           \\ \hline
Pneumonia                  & 6                                                                                                            & 5                                                                                                            \\ \hline
Atelectasis                & 75                                                                                                           & 31                                                                                                           \\ \hline
Pneumothorax               & 1                                                                                                            & 13                                                                                                           \\ \hline
Pleural Effusion           & 11                                                                                                           & 65                                                                                                           \\ \hline
Pleural Other              & 3                                                                                                            & 15                                                                                                           \\ \hline
Fracture                   & 3                                                                                                            & 21                                                                                                           \\ \hline
Support Devices            & 53                                                                                                           & 13                                                                                                           \\ \hline
\end{tabular}%
}
\end{table}

\subsection{Are there significant relationships between conditions labeled from reports and conditions labeled from images?}
To determine whether there are significant relationships between conditions labeled from reports and conditions labeled from images, we learn a mapping from the output of radiologists labeling reports to the output of radiologists labeling images. We then analyze the significant relationships implied by this mapping from a clinical perspective. 

\subsubsection{Method}
We train logistic regression models to map the radiologist report labels for all conditions to the image ground truth for each of the evaluation conditions. We quantitatively measure the relationship between the radiologist report labels and the image ground truth by obtaining odds ratios from the coefficients of these logistic regression models. We review the odds ratios from these models with a board-certified radiologist to understand how particular radiologist report labels might clinically change the odds of image labels.
 

\subsubsection{Training details} We one-hot encode the radiologist report labels and provide these binary variables as inputs to a logistic regression model. For example, the "Atelectasis Positive" variable is 1 if the radiologist labels Atelectasis as positive on the report and 0 otherwise. Similarly, the "Atelectasis Negative" variable is 1 if the radiologist labels Atelectasis as negative on the report and 0 otherwise. The same logic applies to the "Atelectasis Uncertain" variable as well as the other variables for each condition. We then train the logistic regression model with L1 regularization ($\alpha = 0.5$) on the CheXpert test set using the one-hot encoded radiologist report labels (for all conditions) as input and the image ground truth for a condition as output. In total, we train different logistic regression models to map the radiologist report labels to binary image labels for each of the 8 evaluation conditions. We compute odds ratios by exponentiating the coefficients of the logistic regression models.

\subsubsection{Results} 
After training the logistic regression models, we find that particular radiology report labels increased (or decreased) the odds of particular conditions in an image with statistical significance (\textit{P} <  0.05). As expected, we find that radiology report labels associated with a condition increase the odds of that same condition in the image; for example, a Cardiomegaly positive report label increases the odds of Cardiomegaly in the image. We also find that the regression model corrects for label hierarchy. A Cardiomegaly positive report label increases the odds of Enlarged Cardiomediastinum (the parent of Cardiomegaly) on the image by 9.6 times. We similarly observe the model correcting for the label hierarchy of Lung Opacity. Radiology report labels of Edema positive, Consolidation positive, and Atelectasis positive, which all correspond to child conditions of Lung Opacity, increase the odds of Lung Opacity. We also find that the model maps particular uncertainties in report labels to the presence of a condition in the image. For example, Atelectasis uncertain report labels and Edema uncertain report labels increase the odds of Lung Opacity by 2.9 and 7.9 times respectively. 

\begin{figure*}
  \includegraphics[width=\textwidth]{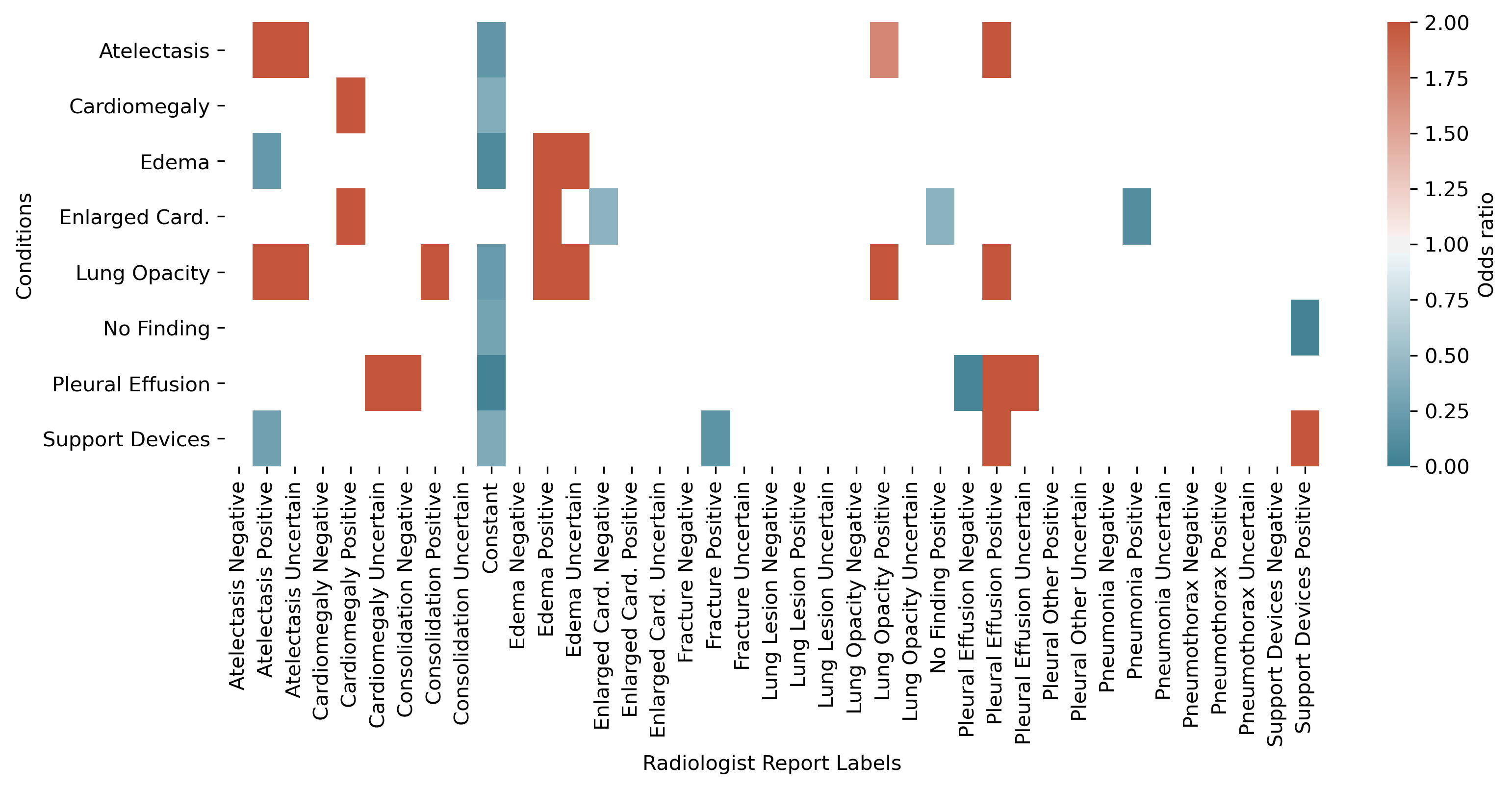}
  \caption{Odds ratios for radiologist report labels as factors for the presence of a condition in the X-ray image. We map the radiologist report labels across all conditions to the image ground truth using a logistic regression model. We obtain odds ratios for the input variables, which are the one-hot encoded radiologist report labels, and only display odds ratios for which the corresponding \textit{P} value (two-sided \textit{t} test) is less than 0.05.}
  \label{fig:odds_ratios}
\end{figure*}

Next, we find that the model maps positive report labels to the presence of other conditions in the image. A Pleural Effusion positive report label increases the odds of Lung Opacity by 4.4 times. We hypothesize that this results from co-occurrence between Pleural Effusion and child conditions of Lung Opacity such as Atelectasis and Edema. Pleural effusion physiologically often leads to adjacent lung collapse, atelectasis, and is often seen in physiologic fluid overload conditions, edema. We find that an Atelectasis positive report label decreases the odds of Support Devices in the image by 0.28 times. On the patient population who have support devices, many of whom are in an Intensive Care Unit (ICU) setting, it is not clinically useful for radiologists to comment on the presence of atelectasis on reports, as they would rather focus on more clinically relevant changes. This may explain the mechanism by which the presence of atelectasis in a report signals that there are no support devices in the image. We find that a Fracture positive report label decreases the odds of Support Devices by 0.17 times. We hypothesize that this results from a negative co-occurrence between Fractures and Support Devices, as the two observations select for different patient populations: X-rays for fractures are often done in the Emergency Department (ED) or other outpatient settings rather than the ICU setting. We find that an Edema positive report label increases the odds of Enlarged Cardiomediastinum on the image by 2.1 times. This may be explained by the fact that Edema and Enlarged Cardiomediastinum often co-occur in a clinical setting, as they can both be caused by congestive heart failure. Lastly, we find that a Support Devices positive report label decreases the odds of No Finding in the image by 0.03 times. This may be explained by the fact that patients with support devices are usually in the ICU setting and sick with other pathologies. We visualize these statistically significant odds ratios for each type of radiologist report label (such as "Atelectasis Negative") as a factor for the presence of an evaluation condition in the X-ray image in Figure \ref{fig:odds_ratios}.

\begin{table}[t!]
\centering
\caption{\label{tab:zero-one_vs_logreg} F1 scores obtained by the Zero-One and LogReg baselines, evaluated on the CheXpert test set. The weighted average is weighted by prevalence (n = \# positive).}
\resizebox{\columnwidth}{!}{%
\begin{tabular}{lcc}
\multicolumn{1}{c}{\textbf{\begin{tabular}[c]{@{}c@{}}Condition \\ (n = \# positive)\end{tabular}}} & \multicolumn{1}{l}{\textbf{Zero-One Baseline}} & \multicolumn{1}{l}{\textbf{LogReg Baseline}} \\ \hline
Atelectasis (n=153) & 0.52 & \textbf{0.63} \\ 
Cardiomegaly (n=151) & 0.46 & \textbf{0.56} \\ 
Edema (n=78) & \textbf{0.53} & 0.47 \\ 
Pleural Effusion (n=104) & \textbf{0.65} & \textbf{0.65} \\ 
\begin{tabular}[c]{@{}l@{}}Enlarged Cardiom. (n=253)\end{tabular} & 0.20 & \textbf{0.67} \\ 
Lung Opacity (n=264) & 0.69 & \textbf{0.81} \\ 
Support Devices (n=261) & \textbf{0.85} & 0.84 \\ 
No Finding (n=62) & 0.39 & \textbf{0.55} \\ \hline
Average & 0.54 & \textbf{0.65} \\ 
Weighted Average & 0.56 & \textbf{0.70} \\
\end{tabular}%
}
\end{table}

\begin{table*}[t!]
\centering
\caption{\label{tab:BERT_logreg} F1 scores for BERT+Thresholding and BERT+LogReg trained on the MIMIC-CXR and CheXpert datasets. We refer to the BERT+Thresholding method on the MIMIC-CXR dataset as VisualCheXbert. The models here are evaluated on the CheXpert test set.}
\resizebox{\textwidth}{!}{%
\begin{tabular}{lcc|cc}
\textbf{\begin{tabular}[c]{@{}l@{}}Condition\\ (n = \# positive)\end{tabular}} & \multicolumn{1}{l}{\textbf{\begin{tabular}[c]{@{}l@{}}BERT+Thresholding on \\ MIMIC-CXR, DenseNet \\ Labels\end{tabular}}} & \multicolumn{1}{l|}{\textbf{\begin{tabular}[c]{@{}l@{}}BERT+LogReg on \\ MIMIC-CXR, DenseNet \\ Labels \\ (VisualCheXbert) \end{tabular}}} & \multicolumn{1}{l}{\textbf{\begin{tabular}[c]{@{}l@{}}BERT+Thresholding on \\ CheXpert, DenseNet \\ Labels\end{tabular}}} & \multicolumn{1}{l}{\textbf{\begin{tabular}[c]{@{}l@{}}BERT+LogReg on \\ CheXpert, DenseNet \\ Labels\end{tabular}}} \\ \hline
Atelectasis (n=153) & \textbf{0.65} & 0.64 & \textbf{0.67} & 0.66 \\
Cardiomegaly (n=151) & 0.53 & \textbf{0.62} & \textbf{0.61} & \textbf{0.61} \\
Edema (n=78) & \textbf{0.55} & 0.54 & 0.49 & \textbf{0.53} \\
Pleural Effusion (n=104) & 0.64 & \textbf{0.65} & 0.57 & \textbf{0.67} \\
\begin{tabular}[c]{@{}l@{}}Enlarged Cardiom. (n=253)\end{tabular} & 0.44 & \textbf{0.73} & 0.60 & \textbf{0.70} \\
Lung Opacity (n=264) & 0.81 & \textbf{0.83} & 0.70 & \textbf{0.83} \\
Support Devices (n=261) & 0.85 & \textbf{0.87} & 0.80 & \textbf{0.84} \\
No Finding (n=62) & 0.44 & \textbf{0.54} & 0.46 & \textbf{0.52} \\ \hline
Average & 0.61 & \textbf{0.68} & 0.61 & \textbf{0.67} \\
Weighted Average & 0.65 & \textbf{0.73} & 0.65 & \textbf{0.72}
\end{tabular}%
}
\end{table*}

\subsection{Can we naively map labels obtained from reports to X-ray image labels?}
We map the output of an automated radiology report labeler to X-ray image labels using simple uncertainty handling strategies. 

\subsubsection{Method}
For a baseline approach, we naively map labels obtained from running the CheXpert labeler on the radiology report impressions to X-ray image labels. The CheXpert labeler is an automatic, rules-based radiology report labeler \cite{irvin2019chexpert}. The labels produced by the CheXpert labeler include 4 classes per medical condition (positive, negative, uncertain, and blank). Since the image ground truth only has positive or negative labels per condition, we must map the labels produced by the CheXpert labeler to binary labels. We map the blank labels produced by the CheXpert labeler to negative labels. We do not change the positive and negative labels produced by the CheXpert labeler. To handle the uncertain labels, we use the two common uncertainty handling strategies in \citet{irvin2019chexpert}: we map the uncertain labels to either all negative labels (zeros-uncertainty handling strategy) or all positive labels (ones-uncertainty handling strategy). We record the F1 score from the better performing strategy on the CheXpert test set, using as ground truth the labels provided by radiologists labeling X-ray images (image ground truth). We refer to this method as the \textit{Zero-One Baseline}. Since we only report the maximum of the zeros-uncertainty handling strategy and the ones-uncertainty handling strategy, the F1 scores for the Zero-One Baseline represent the most optimistic global mapping of the uncertainty labels for this method.

\subsubsection{Results}
We find that the average and weighted average F1 scores across the evaluation conditions for the Zero-One Baseline are 0.54 and 0.56 respectively, which are in between the average / weighted average ``Low F1'' and ``High F1'' scores for radiologists labeling reports (see Table \ref{tab:report_image_agreement}). This indicates that the Zero-One Baseline is not strictly better or worse than radiologists labeling reports, who we previously show to have poor agreement with radiologists labeling images. The Zero-One Baseline F1 scores for Atelectasis, Cardiomegaly, Edema, Pleural Effusion, and Enlarged Cardiomediastinum are 0.52, 0.46, 0.53, 0.65, and 0.20 respectively, which are all between the respective ``Low F1'' and ``High F1'' scores for radiologists labeling reports. The Zero-One Baseline F1 scores for Lung Opacity and No Finding are 0.69 and 0.39 respectively, which are slightly higher ($\sim 0.01$ difference) than the respective ``High F1'' scores for radiologists labeling reports. Similarly, the Zero-One Baseline F1 score for Support Devices is 0.39, which is slightly lower ($\sim 0.01$ difference) than the ``Low F1'' Support Devices score for radiologists labeling reports. The F1 scores for the Zero-One Baseline across the evaluation conditions are shown in Table \ref{tab:zero-one_vs_logreg}.

\subsection{Can we learn to map labels obtained from reports to X-ray image labels?}
We map the output of an automated radiology report labeler to X-ray image labels, similarly to how we previously map the output of radiologists labeling reports to the output of radiologists labeling images. Previous work by \citet{dunnmon2019crossmodal} showed that labels obtained from noisy labeling functions on radiology reports can be mapped to labels that are of similar quality to image labels produced by radiologists for the simpler task of classifying X-rays as normal or abnormal.

\subsubsection{Method}
This approach, motivated by a prior experiment in which we map radiologist report labels to image labels, improves upon the naive uncertainty mapping strategy used in the Zero-One Baseline. As before, we obtain report labels by running the CheXpert labeler on radiology report impressions. For each of the evaluation conditions, we train a logistic regression model that maps the CheXpert labeler’s output on a radiology report impression to a positive or negative label for the target condition. This approach makes use of the automated report labels for all 14 conditions to predict the label for each target condition. We refer to this approach as the \textit{LogReg Baseline}.

\subsubsection{Training details} 
We one-hot encode the report labels outputted by the CheXpert labeler and provide these binary variables as inputs to a logistic regression model. We train a logistic regression model with $L2$ regularization ($C=1.0$) and a max iteration of $500$ using the one-hot encoded report labels (for all conditions) as input and the image ground truth for a condition as output. The class weights are the inverse prevalence of the respective class in the training set. We use a leave-one-out cross-validation strategy to train and validate the logistic regression model on the CheXpert test dataset. For each of the 8 evaluation conditions, we train different logistic regression models to map the labels produced by the CheXpert labeler to binary image labels.

\subsubsection{Results}
We find that the LogReg Baseline approach improves upon the Zero-One Baseline for most conditions. Compared to the Zero-One Baseline, the LogReg Baseline increases the average F1 score from 0.54 to 0.65 and the weighted average F1 score from 0.56 to 0.70. The LogReg Baseline increases the F1 score compared to the Zero-One Baseline from 0.52 to 0.63 for Atelectasis, 0.46 to 0.56 for Cardiomegaly, 0.20 to 0.67 for Enlarged Cardiomediastinum, 0.69 to 0.81 for Lung Opacity, and 0.39 to 0.55 for No Finding. However, the LogReg Baseline decreases the F1 scores compared to the Zero-One Baseline from 0.53 to 0.47 for Edema and 0.85 to 0.84 for Support Devices. For Pleural Effusion, both the LogReg Baseline and the Zero-One Baseline have an F1 score of 0.65. Although the LogReg Baseline is not better than the Zero-One Baseline for all conditions, these results suggest that a learned mapping from radiologist report labels to X-ray image labels can outperform naively mapping all uncertain labels to positive or negative for most conditions. The F1 scores obtained by the LogReg Baseline, along with head-to-head comparisons to the Zero-One Baseline, are shown in Table \ref{tab:zero-one_vs_logreg}.

\begin{table*}[t!]
\centering
\caption{\label{tab:improvement} Improvement in F1 score obtained by VisualCheXbert, evaluated on the CheXpert test set and reported with 95\% confidence intervals. The left-most column shows the improvement over the Zero-One Baseline. The middle column shows the improvement over the radiologist report labels with uncertains mapped to the image ground truth label. The right-most column shows the improvement over the radiologist report labels with uncertains mapped to the opposite of image ground truth label.}
\resizebox{1.5\columnwidth}{!}{%
\begin{tabular}{lccc}
\textbf{\begin{tabular}[c]{@{}l@{}}Condition \\ (n = \# positive)\end{tabular}} & \multicolumn{1}{l}{\textbf{\begin{tabular}[c]{@{}l@{}}Improvement over\\ Zero-One Baseline\end{tabular}}} & \multicolumn{1}{l}{\textbf{\begin{tabular}[c]{@{}l@{}}Improvement over\\ Higher Radiologist\\ Score\end{tabular}}} & \multicolumn{1}{l}{\textbf{\begin{tabular}[c]{@{}l@{}}Improvement over\\ Lower Radiologist\\ Score\end{tabular}}} \\ \hline
Atelectasis (n=153) & 0.12 (0.04, 0.20) & 0.04 (-0.04, 0.12) & 0.41 (0.32, 0.49) \\
Cardiomegaly (n=151) & 0.16 (0.07, 0.25) & 0.15 (0.07, 0.25) & 0.20 (0.11, 0.28) \\
Edema (n=78) & 0.01 (-0.05, 0.07) & -0.04 (-0.09, 0.02) & 0.09 (0.02, 0.17) \\
Pleural Effusion (n=104) & -0.01 (-0.04, 0.03) & -0.06 (-0.10, -0.02) & 0.01 (-0.03, 0.05) \\
Enlarged Cardiom. (n=253) & 0.53 (0.46, 0.60) & 0.52 (0.44, 0.60) & 0.64 (0.57, 0.71) \\
Lung Opacity (n=264) & 0.14 (0.09, 0.20) & 0.15 (0.09, 0.20) & 0.15 (0.10, 0.20) \\
Support Devices (n=261) & 0.02 (-0.01, 0.06) & 0.01 (-0.02, 0.04) & 0.01 (-0.02, 0.04) \\
No Finding (n=62) & 0.15 (0.05, 0.26) & 0.16 (0.05, 0.28) & 0.16 (0.05, 0.28) \\ \hline
Average & 0.14 (0.12, 0.17) & 0.12 (0.09, 0.15) & 0.21 (0.18, 0.24) \\
Weighted Average & 0.17 (0.15, 0.20) & 0.15 (0.13, 0.18) & 0.24 (0.21, 0.26)
\end{tabular}%
}
\end{table*}

\subsection{Can we learn to map the text reports directly to the X-ray image labels?}
Previously, we mapped the output of an existing automated report labeler, which takes text reports as input, to X-ray image labels. We now map the textual radiology report directly to the X-ray image labels.
\label{deep_learning_approach}


\subsubsection{Method}
\label{BERT_model_method}
 We develop a deep learning model that maps a radiology report directly to the corresponding X-ray image labels.  

 Since it is too expensive to obtain labels from radiologists for hundreds of thousands of X-ray images to supervise our model, we instead train a single DenseNet model \cite{huang2018densely} to detect medical conditions from chest X-ray images, as is described by \citet{irvin2019chexpert}, and we use this computer vision model as a proxy for a radiologist labeling an X-ray image. We use the DenseNet model to output probabilities for each of the 14 conditions for all X-rays in the MIMIC-CXR dataset and the CheXpert training dataset. To obtain the output of the vision model on the MIMIC-CXR dataset, we train the DenseNet on the CheXpert training dataset. Similarly, to obtain the output of the vision model on the CheXpert training dataset, we train the DenseNet on the MIMIC-CXR dataset. We find that the DenseNet trained on the CheXpert training set has an AUROC of 0.875 on the CheXpert test set across all conditions, and the DenseNet trained on the MIMIC-CXR dataset has an AUROC of 0.883 on the CheXpert test set across all conditions.

We then use the probabilities outputted from these computer vision models as ground truth to fine-tune a BERT-base model. We train one BERT model using the MIMIC-CXR dataset and one using the CheXpert training dataset. The BERT model takes a tokenized radiology report impression from the MIMIC-CXR or CheXpert dataset as input and is trained to output the labels produced by the DenseNet model. We feed the BERT model's output corresponding to the \textit{[CLS]} token into linear heads (one head for each medical condition) to produce scores for each medical condition. We use the cross-entropy loss to fine-tune BERT. The BERT model is initialized with biomedically pretrained weights produced by \citet{peng2019transfer}. This model training process is shown in Figure \ref{fig:teaser}.

After training the BERT model, we map the outputs of BERT, which are probabilities, to positive or negative labels for each condition. To do so, we try two different methods. Our first method uses optimal probability thresholds to convert the BERT outputs to binary labels. We calculate optimal thresholds by finding the threshold for each condition that maximizes Youden's index \cite{youden1950index} (the sum of sensitivity and specificity minus one) on the CheXpert validation dataset. We refer to this approach as \textit{BERT+Thresholding}. Our second method trains a logistic regression model to map the output of BERT across all 14 conditions to a positive or negative label for the target condition. We refer to this approach as \textit{BERT+LogReg}. Ultimately, we develop four different models by using both methods on outputs from a BERT model trained on the MIMIC-CXR dataset and a BERT model trained on the CheXpert training dataset. The four resulting models are called BERT+Thresholding on MIMIC-CXR, BERT+LogReg on MIMIC-CXR, BERT+Thresholding on CheXpert, and BERT+LogReg on CheXpert. We refer to the BERT+LogReg model trained on the MIMIC-CXR dataset with labels provided by the DenseNet model, which is our best performing approach, as \textit{VisualCheXbert}.

\subsubsection{Training details}
\label{BERT_training_details}
We train the BERT model on 3 TITAN-XP GPUs using the Adam optimizer \cite{kingma2017adam} with a learning rate of $2 \times 10^{-5}$, following \citet{devlin2019bert} for fine-tuning tasks. We use a random 85\%-15\% training-validation split, as in \citet{smit2020chexbert}. The BERT model is trained until convergence. We use a batch size of 18 radiology report impressions. For the BERT+LogReg approach, the logistic regression model uses $L2$ regularization ($C = 1.0$) and a max iteration of 500. Similar to the LogReg Baseline, the class weights are the inverse prevalence of the respective class in the training set, and we use a leave-one-out cross-validation strategy to train and test the logistic regression model on the CheXpert test dataset. We train different logistic regression models to map the probabilities outputted by the BERT model to the binary image labels for each of the 8 evaluation conditions.

\subsubsection{Results}
We compare the performance of the different BERT approaches on the CheXpert test set.  First, we find that on most conditions, BERT+LogReg outperforms BERT+Thresholding. This finding holds true on both the CheXpert and MIMIC-CXR datasets. Second, we find that despite being trained on datasets from different institutions, the models trained on MIMIC-CXR and CheXpert datasets perform similarly. This indicates that the BERT model trained on radiology report impressions from the MIMIC-CXR distribution (Beth Israel Deaconess Medical Center Emergency Department between 2011–2016) \cite{johnson2019mimiccxrjpg} can perform as well as a model trained on radiology report impressions from the CheXpert distribution (Stanford Hospital between 2002-2017) \cite{irvin2019chexpert}, even when both models are evaluated on a test set from the CheXpert distribution. Since we obtain a slightly higher average and weighted average F1 using the MIMIC-CXR dataset, we use BERT trained on MIMIC-CXR in our final approach called VisualCheXbert. The performance of the BERT approaches is shown in Table \ref{tab:BERT_logreg}.

Next, we compare VisualCheXbert to the Zero-One Baseline. When comparing VisualCheXbert to the Zero-One Baseline as well as the higher and lower scores of radiologists labeling reports described below, we report the improvements by computing the paired differences in F1 scores on 1000 bootstrap replicates and providing the mean difference along with a 95\% two-sided confidence interval \cite{efron1986bootstrap}. Overall, VisualCheXbert improves the average F1 and weighted average F1 over the Zero-One Baseline with statistical significance, increasing the average F1 score by 0.14 (95\% CI 0.12, 0.17) and the weighted average F1 score by 0.17 (95\% CI 0.15, 0.20). We find that VisualCheXbert obtains a statistically significant improvement over the Zero-One Baseline on most conditions. VisualCheXbert increases the F1 score on Enlarged Cardiomediastinum, Cardiomegaly, No Finding, Lung Opacity, and Atelectasis compared to the Zero-One Baseline by 0.53 (95\% CI 0.46, 0.60), 0.16 (95\% CI 0.07, 0.25), 0.15 (95\% CI 0.05, 0.26), 0.14 (95\% CI 0.09, 0.20), and 0.12 (95\% CI 0.04, 0.20), respectively. VisualCheXbert obtains similar performance (no statistically significant difference) to the Zero-One Baseline on the rest of the conditions, which are Edema, Pleural Effusion, and Support Devices, with improvements of 0.01 (95\% CI -0.05, 0.07), -0.01 (95\% CI -0.04, 0.03), and 0.02 (95\% CI -0.01, 0.06), respectively.

Lastly, we compare the F1 scores for VisualCheXbert to the higher and lower scores of radiologists labeling reports. The higher scores for radiologists labeling reports are obtained by mapping the uncertain radiologist report labels to the image ground truth label, while the lower scores for radiologists labeling reports are obtained by mapping the uncertain radiologist report labels to the opposite of the ground truth. Overall, VisualCheXbert obtains a statistically significant improvement over both the higher and lower radiologist scores, increasing the average F1 score by 0.12 (95\% CI 0.09, 0.15) over the higher radiologist score and 0.21 (95\% CI 0.18, 0.24) over the lower radiologist score and increasing the weighted average F1 score by 0.15 (95\% CI 0.13, 0.18) over the higher radiologist score and 0.24 (95\% CI 0.21, 0.26) over the lower radiologist score. Statistically significant improvements over the higher radiologist score are observed for Cardiomegaly (0.15 [95\% CI 0.07, 0.25]), Enlarged Cardiomediastinum (0.52 [95\% CI 0.44, 0.60]), Lung Opacity (0.15 [95\% CI 0.09, 0.20]), and No Finding (0.16 [95\% CI 0.05, 0.28]). VisualCheXbert performs similarly (no statistically significant difference) to the higher radiologist score on Atelectasis (0.04 [95\% CI -0.04, 0.12]), Edema (-0.04 [95\% CI -0.09, 0.02]), and Support Devices (0.01 [95\% CI -0.02, 0.04]). VisualCheXbert performs slightly worse than the higher radiologist score on one condition, which is Pleural Effusion (-0.06 [95\% CI -0.10, -0.02]). VisualCheXbert observes considerable, statistically significant improvements compared to the lower radiologist score on all but two conditions. There is no statistically significant difference between VisualCheXbert and the lower radiologist score on these two conditions, which are Pleural Effusion (0.01 [95\% CI -0.03, 0.05]) and Support Devices (0.01 [95\% CI -0.02, 0.04]). We show the improvements obtained by VisualCheXbert over the Zero-One Baseline and the improvements over radiologists labeling reports in Table \ref{tab:improvement}.

\section{Limitations}
Our work has the following limitations. First, our study only made use of the \textit{Impression} section of the radiology reports, which is a summary of the radiology report. Prior work regarding automated chest X-ray labeling has also extensively used the impression section in radiology reports \cite{irvin2019chexpert, johnson2019mimiccxrjpg, wang2017chestx}. However, conditions are sometimes mentioned in the \textit{Findings} section of the report but not in the \textit{Impression} section. As a result, negative and blank labels are more frequent when using the \textit{Impression} section, and this could increase the disparity between labels extracted from the impression and the corresponding chest X-ray image labels. Second, the VisualCheXbert model has a maximum input size of 512 tokens. In practice, only 3 of the report impressions in the entire CheXpert dataset were longer than this limit. Third, the CheXpert test set, on which we evaluated our models, consists of 500 radiology studies and is therefore limited in size. As a result, some of the medical conditions contained very few positive examples; we only evaluated our models on conditions for which at least 10\% of the examples in the CheXpert test set were positive. Using a larger test set would allow evaluation on rarer conditions. Fourth, our models are evaluated on chest X-rays from a single institution. Further evaluation on data from other institutions could be used to evaluate the generalizability of our models.


\section{Conclusion}
We investigate the discrepancy between labels extracted from radiology reports and the X-ray image ground truth labels. We then develop and evaluate methods to address this discrepancy. In our work, we aim to answer the following questions.

\textit{Do radiologists labeling reports agree with radiologists labeling X-ray images?} We find that there is significant disagreement between radiologists labeling reports and radiologists labeling images. On the CheXpert test set, we observe low Kappa scores for almost all conditions evaluated. The average Kappa across the evaluation conditions is between 0.312 and 0.430. These bounds are based on the most pessimistic mapping and most optimistic mapping of uncertain radiology report labels.

\textit{Why do radiologists labeling reports disagree with radiologists labeling X-ray images?} Upon a board-certified radiologist review of examples of disagreements between radiologists labeling reports and radiologists labeling images, we find four main reasons for disagreement. First, on the CheXpert test set, radiologists labeling reports typically do not mark a parent condition as positive if a child condition is positive. An example of a parent and child condition would be Lung Opacity and Edema, respectively. Second, radiologists labeling reports have access to clinical report history, which biases their diagnoses compared to radiologists labeling images who do not have access to this information. Third, conditions are sometimes reported in the \textit{Findings} section of radiology reports but not the \textit{Impression} section of radiology reports. However, the \textit{Impression} section of radiology reports is commonly used to label reports. This discrepancy can cause radiologists labeling reports to miss pathologies present on the X-ray image. Fourth, labeling images and reports is noisy to a certain extent due to factors such as human mistakes, uncertainty in both reports and images, and different thresholds for diagnosing conditions as positive among radiologists.  

\textit{Are there significant relationships between conditions labeled from reports and conditions labeled from images?} We find many significant relationships between conditions labeled from reports and conditions labeled from images. We report and clinically interpret various radiology report labels that increase (or decrease) the odds of particular conditions in an image with statistical significance (\textit{P}< 0.05). As expected, we find that positive report labels for a condition increase the odds of that condition in an image. We find that positive report labels for children of a condition increase the odds of the parent condition in an image, a phenomenon that is correcting for the label hierarchy. We find that particular uncertain report labels for a condition increase the odds of the condition (and/or its parent condition). We also find that positive report labels for certain conditions increase (or decrease) the odds of other conditions in the image. One example is that a positive Atelectasis report label decreases the odds of Support Devices in the X-ray image by 0.28 times. We explain potential mechanisms by which the presence of a condition in a report signals the presence (or absence) of another condition in the image.

\textit{Can we learn to map radiology reports directly to the X-ray image labels?}
We learn to map a textual radiology report directly to the X-ray image labels. We use a computer vision model trained to detect diseases from chest X-rays as a proxy for a radiologist labeling an X-ray image. Our final model, VisualCheXbert, uses a biomedically-pretrained BERT model that is supervised by the computer vision model. When evaluated on radiologist image labels on the CheXpert test set, VisualCheXbert increases the average F1 score across the evaluation conditions by between 0.12 (95\% CI 0.09, 0.15) and 0.21 (95\% CI 0.18, 0.24) compared to radiologists labeling reports. VisualCheXbert also increases the average F1 score by 0.14 (95\% CI 0.12, 0.17) compared to a common approach that uses a previous rules-based radiology report labeler.

Given the considerable, statistically significant improvement obtained by VisualCheXbert over the approach using an existing radiology report labeler \cite{irvin2019chexpert} when evaluated on the image ground truth, we hypothesize that VisualCheXbert's labels could be used to train better computer vision models for automated chest X-ray diagnosis.

\section{Code Repository}
The code to run our model is available in a public code repository: \url{https://github.com/stanfordmlgroup/VisualCheXbert}.

\begin{acks}
We would like to acknowledge the Stanford Machine Learning Group (\url{stanfordmlgroup.github.io}) and the Stanford Center for Artificial Intelligence in Medicine and Imaging (\url{AIMI.stanford.edu}) for infrastructure support.
\end{acks}

\bibliographystyle{ACM-Reference-Format}
\bibliography{sample-base}






\end{document}